\documentclass[aps,pra,reprint,twocolumn,superscriptaddress,floatfix,showpacs,nofootinbib]{revtex4-1}
\usepackage{epsfig}
\usepackage{amsmath,amssymb,amsfonts}
\usepackage{hyperref}
\usepackage{mathrsfs}
\usepackage{bbm}
\usepackage{slashed}
\usepackage{graphicx}
\usepackage{verbatim}
\usepackage{url}
\usepackage[normalem]{ulem}

\usepackage[usenames]{color}

\newcommand{\be}{\begin{equation}}
\newcommand{\ee}{\end{equation}}

\newcommand{\bi}{\begin{itemize}}
\newcommand{\ei}{\end{itemize}}
\newcommand{\bea}{\begin{eqnarray}}
\newcommand{\eea}{\end{eqnarray}}
\newcommand{\ud}{\mathrm{d}}

\newcommand{\Chris}[1]{{\textcolor{black}{#1}}}  
\newcommand{\Arkady}[1]{{\textcolor{black}{#1}}}  
\newcommand{\RefB}[1]{{\textcolor{black}{#1}}}  
\definecolor{darkgreen}{rgb}{0.1,0.7,0}
\definecolor{bensblue}{rgb}{0.1,0.1,0.9}

\usepackage[T1]{fontenc} \usepackage[latin1]{inputenc}

\begin{document}
\title{Narrowing of the emission angle in high-intensity Compton scattering}

\author{C.~N.~Harvey}
\email[]{christopher.harvey@chalmers.se}
\affiliation{Department of Applied Physics, Chalmers University of Technology, SE-41296 Gothenburg, Sweden}
\author{A.~Gonoskov}
\email[]{arkady.gonoskov@chalmers.se}
\affiliation{Department of Applied Physics, Chalmers University of Technology, SE-41296 Gothenburg, Sweden}
\affiliation{Institute of Applied Physics, Russian Academy of Sciences, Nizhny Novgorod 603950, Russia}
\affiliation{University of Nizhny Novgorod, Nizhny Novgorod 603950, Russia}
\author{M.~Marklund}
\email[]{mattias.marklund@chalmers.se}
\affiliation{Department of Applied Physics, Chalmers University of Technology, SE-41296 Gothenburg, Sweden}
\author{E.~Wallin}
\email[]{erik.wallin@physics.umu.se}
\affiliation{Department of Physics, Ume\aa\ University,  SE-90187 Ume\aa, Sweden}

\begin{abstract}
We consider the emission spectrum of high-energy electrons in an intense laser field.  At high intensities ($a_0\sim200$) we find that the QED theory predicts a narrower angular spread of emissions than the classical theory.  This is due to the classical theory overestimating the energy loss of the particles, resulting in them becoming more susceptible to reflection in the laser pulse.
\end{abstract}
\maketitle

\section{Introduction}
Since the discovery of chirped pulse amplification \cite{Strickland1985219} the  powers and intensities of state of the art laser facilities have been exponentially increasing \cite{PhysRevSTAB.5.031301}, the current record of $2\times 10^{22}$ W cm$^{-2}$ having been set in 2008 \cite{Yanovsky:2008}.  With the advent of various new facilities over the next few years, such as the Vulcan 20 PW upgrade \cite{Vulcan}, the Extreme Light Infrastructure (ELI) Facility \cite{ELI} and the XCELS project \cite{XCELS}, this trend is expected to continue for the foreseeable future.  The availability of such technology has driven a large field of research in the topic of nonlinear Thomson and Compton scattering, the understanding of which is important from a fundamental physics perspective \cite{Heinzl:2011ur, DiPiazza:2012RevModPhys}.  Additionally, and at least as importantly, the process produces high-energy, tuneable $\gamma$-ray beams, which are important for fundamental research \cite{Wu}, as well as more practical applications such as cancer radiotherapy \cite{Lawrence} and the radiography of dense objects \cite{PhysRevLett.94.025003}.  Recent experiments \cite{2011NatPh...7..867C, PhysRevLett.113.224801} have been pushing the limits of peak energies and brilliances, taking us towards the regime where radiation reaction and QED effects will start to come into play \cite{Vranic:2014, PhysRevX.2.041004,Green:2014}.
(For related studies of high energy electrons in orientated crystals see, e.g., \cite{baier1998electromagnetic,Baurichter:1997,Kirsebom:2001}.)

In this article we study nonlinear Thomson and Compton scattering at ultra high intensities, assessing the impact of classical radiation reaction and QED effects on the properties of the emitted photon spectra.  We find that, because the classical theory overestimates the radiative energy loss, it predicts a broader angular spread of emissions than the QED theory.

\section{Theory}
We consider the case of an electron in a head on collision with a laser pulse.  We adopt natural units where $\hbar=c=1$. To begin with we will take our laser to be a plane wave field propagating in the $z$-direction described by the null wave vector $k^{\mu}=\omega_0(1,0,0,1)$, with central frequency $\omega_0$.  We assume the field to be polarised in the perpendicular ($x$) direction and therefore introduce the polarisation vector $\epsilon=(0,1,0,0)$.  We define the dimensionless intensity in the usual manner, 
$a_0=eE/\omega_0 m $,
where $E$ is the {\it peak} magnitude of the electrical field strength.  The electromagnetic field tensor of the wave is taken to depend arbitrarily on the phase $\phi\equiv k\cdot x=\omega(t-z)$; $F^{\mu\nu}(\phi)=a_0f(\phi)f^{\mu\nu}$, where $f^{\mu\nu}=(k^\mu\epsilon^\nu-k^\nu\epsilon^\mu)/\omega$ and $f(\phi)$ is a function describing the pulse.  Additionally, to aid future discussion, we also define the time-dependent intensity $a(\phi)=e f(\phi)E  /\omega m$, which is equal to $a_0$ at the pulse peak. 

The emission spectrum from a particle in the field can be decomposed into a sum of harmonics, corresponding to multiples of the laser frequency.
In the quantum description these correspond to the number of laser photons involved in the scattering process \cite{Ritus:1985}.
During each scattering process, the electron will absorb an integer number $n$ of laser photons (each of momentum $k$) before emitting a photon of momentum $k^\prime$.
From conservation of momentum arguments it can be shown that the frequency of the scattered photon is given by \cite{Harvey:2009}
\begin{eqnarray}
\omega_{n}^{\prime}=\frac{n\omega_0}{1+j_n (1-\cos{\theta})},\label{nuprime}
\end{eqnarray}
where 
\begin{eqnarray}
j_n=\frac{n\omega_0/m -\gamma\beta +a_0^2\gamma (1-\beta )/2}{\gamma (1+\beta )}.\label{jn}
\end{eqnarray}
It can be seen that when $j_n <0$ the maximum emission frequency occurs when the photons are backscattered
($\theta =180^\circ$).  Conversely, when $j_n >0$ the maximum frequency occurs for forward
scattering ($\theta =0$).  Note that the support for a given harmonic depends on the harmonic number $n$ (and that the amplitude of a given harmonic will be determined by the cross sections provided in the references).
For high intensities the spectrum will be composed of a very large number of harmonics, with the spectrum decaying after the harmonic with number $n\sim 3 a_0^3/2$ \cite{Esarey:1993}.

\begin{figure}
\includegraphics[width=1.0\columnwidth,]{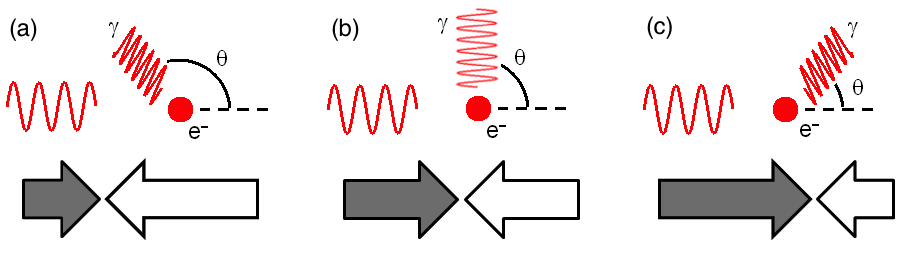}
\caption{Diagrams showing the relationship between the laser momentum and the peak scattering angle of the emitted photons.  (a)   $a_0 \ll \gamma$, the peak emissions will be forward scattered relative to the laser axis.  (b) $a_0\sim 2\gamma$ defines the `center-of-mass' frame for the collision where the peak emissions will be at $90^\circ$.  (c) $a_0 \gg \gamma$ the peak emissions will be back scattered relative to the laser axis.  See Ref.~\cite{Harvey:2009} for further details.  \label{fig:centre_of_mass} }
\end{figure}

Now let us consider the angular directions of the emissions.  From expressions (\ref{nuprime}) and (\ref{jn}) it can be seen that the angular range of each harmonic will depend on both $a_0$ and the $\gamma$-factor of the electron.  It can be shown that the angle at which the emissions peak depends on the ratio of the sum of the laser photon momenta to the electron momenta (see \cite{Harvey:2009} for further details).
In the case where \RefB{the electron} momentum is greater than the sum of the laser photon momenta for all the photons involved in even the highest order scattering processes, i.e. $a_0 \ll \gamma$, the peak emissions will be forward scattered relative to the laser axis.  In the opposite case, $a_0 \gg \gamma$ they will be backscattered, and the case $a_0\sim 2\gamma$ defines the `center-of-mass' frame for the collision (see \cite{Harvey:2009, Harvey:2012mass_shift}) where the peak emissions will be at $90^\circ$.  This is demonstrated in Fig.~\ref{fig:centre_of_mass}.

\begin{figure}
\includegraphics[width=0.8\columnwidth,]{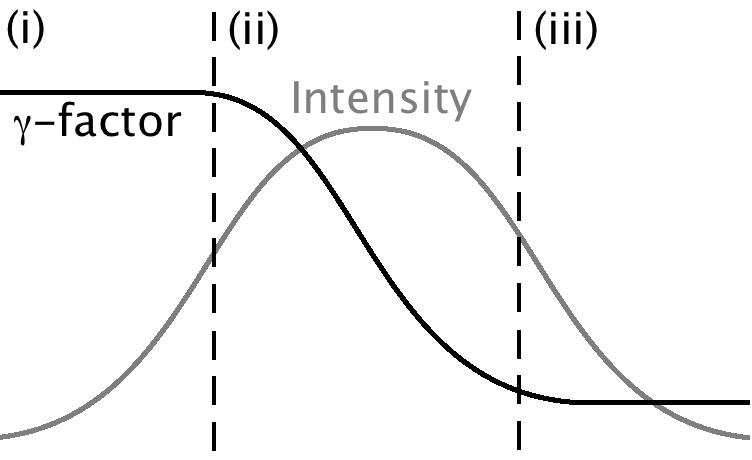}
\caption{Diagram showing how the $\gamma$-factor of the particle decreases due to RR loses throughout its interaction with the laser pulse.  The dynamics can be roughly divided into three regimes: (i) Initially $\gamma\gg a(\phi)$, (ii) as the particle approaches the peak field it loses energy due to RR, resulting in $\gamma\sim a(\phi)$, (iii) after the peak field there will be a period where $\gamma < a(\phi)$.\label{fig:electron_energy} }
\end{figure}

When the background field is of high intensity the radiation emissions from the particle will be so strong that the resulting energy loss will begin to effect its motion \cite{Harvey:2012gaussian}.  This radiation reaction (RR) effect will cause  the $\gamma$-factor of the particle to decrease during the interaction with the laser field.  This will mean that in order to estimate the peak emission angle we must consider the value of the $\gamma$-factor when the particle is in the peak of the laser field (since the highest intensity region will produce the strongest emission signal strength and therefore dominate the total spectrum), rather than the initial value.  \RefB{This is illustrated} in Fig.~\ref{fig:electron_energy} where we show how the ratio of laser intensity to particle $\gamma$-factor changes with time.

If we wish to consider quantum effects then it is instructive to introduce the dimensionless and invariant  `quantum efficiency' parameter \RefB{$\chi_e\equiv \sqrt{p^\mu F^{2}_{\mu\nu}p^\nu}/m^2\sim\gamma E/E_{\rm cr}$}, where $E_{\rm cr}=1.3\times10^{16}$\,Vcm$^{-1}$ is the QED `critical' field (`Sauter-Schwinger' field)~\cite{QEDcriticalfield1,*QEDcriticalfield2,*QEDcriticalfield3}.  
This can be interpreted as the work done on the electron by the laser field over the distance of a Compton wavelength.  
In the regime $\chi_e\gtrsim1$ quantum effects, including pair production from the emitted photons, will dominate.  
For the purposes of this study we restrict ourselves to the regime where $a_0$, $\gamma\gg 1$, such that quantum effects play a role in the Compton spectra, but have $\chi_e\lesssim 1$, so that pair production can be neglected. 

\section{Numerical techniques}
To generate the emission spectra we use the single particle code SIMLA \cite{Green:2014kfa}, which runs in both classical and QED modes.  
In the classical case we propagate the particle through the fields using the Landau Lifshitz (LL) equation, which takes into accounts RR effects via the inclusion of some correctional terms to the Lorentz force equation~\cite{LLII}
\begin{equation}
\begin{split}
 \dot{u}^\mu = & F^{\mu\nu} u_\nu + r_0 \big(
  \dot{F}^{\mu\nu} u_\nu +\\
 & F^{\mu\alpha}F_\alpha^{\;\;\nu} u_\nu -  u_\alpha
  F^{\alpha\nu}F_\nu^{\;\;\beta} u_\beta \, u^\mu \big),\label{LL1}
  \end{split}
\end{equation}
where $r_0\equiv (2/3)e^2/4\pi m$ is the classical electron radius, and the overdot denotes differentiation with respect to proper time.  We note that there are numerous alternative equations in the literature (for an overview see Refs.~\cite{Burton:2014, Vranic:2015} and for further discussions of the LL equation see Refs.~\cite{Harvey:2011dp,Hadad:2010}).  However, the LL equation has, along with some others, recently been shown to be consistent with QED to the order of the fine-structure constant $\alpha$ \cite{Ilderton:2013tb}.

The resulting classical emission spectra are calculated using a novel Monte Carlo method introduced in \cite{Wallin:2014moa}.
The method is simple and computationally efficient.
As the particles in our simulation are ultra-relativistic the radiation due to transverse acceleration is dominant, since this is a factor $\gamma^2$ larger than that due to longitudinal acceleration \cite{jackson_classical_1999}.
Since the acceleration and velocity of the particle are perpendicular, we can approximate the radiation as synchrotron radiation.
In our method we calculate the \emph{effective magnetic field}, $H_{\text{eff}}$, acting on the particle over each timestep. This is the magnetic field which would cause the same acceleration as the electric and magnetic fields together.
The typical frequency of synchrotron emission is then given by $\omega_c=3eH_{\text{eff}}\gamma^2/2m$. 
For an ultra-relativistic particle in an external, homogenous magnetic field, 
\Chris{the classical radiation cross section can be expressed in terms of the intensity given by \cite{jackson_classical_1999}
\begin{equation}
	\frac{ \partial \Gamma_\textrm{cl}}{\partial \omega^\prime} = \frac{1}{\omega^\prime} \frac{ \partial I}{\partial \omega^\prime} = \frac{\sqrt{3}}{2 \pi} \frac{e^3 H_\textrm{eff}}{\omega^\prime m} F_1(\omega^\prime/\omega_c),\label{Erikclassical}
\end{equation}
where}
\RefB{$F_1(\xi) = \xi \int_{\xi}^{\infty} K_{5/3}(\xi^\prime) \mathrm{d}\xi^\prime$} is the first synchrotron function.
\Chris{(We note that (\ref{Erikclassical}) is integrable in the limit $\omega^\prime\rightarrow 0$ and therefore the expression is well-defined. For further details see Ref.~\cite{Khokonov:2002}.)}
At each timestep the quantity $\omega_c$ is calculated and a Monte-Carlo method is used to sample from the spectra.
The direction of the emission is taken to be that of the particle velocity, a good approximation for the ultra-relativistic case \cite{jackson_classical_1999}.

In the QED case our code works as follows.  
\Chris{
We begin by introducing the (quantum) synchrotron parameter\footnote{Note that in the classical regime the energy of the emitted radiation is typically much lower than the electron energy $m\gamma$ (since the radiation is emitted continuously rather than as discrete emissions) and so we can approximate $\xi\approx (2/3\chi_e)(\omega^\prime/\gamma)$. Further approximating $\chi_e\approx \gamma H_\textrm{eff} e/m^2$, one can see that $\xi\approx\omega^\prime/\omega_c$, the argument of our classical expression (\ref{Erikclassical}).} 
in terms of the frequency of the emitted radiation $\omega^\prime$
\begin{equation}
\xi=\frac{2}{3\chi_e}\frac{\omega^\prime}{\gamma m-\omega^\prime}. 
\end{equation}
In the case of high energy emissions, $\xi\gg 1$, it can be shown that the photon coherence length goes like \cite{Khokonov:2010}
\begin{eqnarray}
l_f\approx \sqrt{\frac{2}{3}}\frac{\lambda}{\pi a_0}\sqrt{\frac{1}{\xi}},
\end{eqnarray}
where $\lambda=2\pi/\omega_0$ is the laser wavelength.  Thus in the high-intensity limit $a_0\gg 1$ the size of the radiation formation region is much smaller than the laser wavelength~\cite{Ritus:1985} and so the laser background field can be approximated as locally constant and crossed~\cite{Nikishov:1964zza}.  We are then able to determine the probability of photon emission using the expression for the constant crossed field rate $\Gamma_{\textrm{q}}$ per unit time,}
\begin{equation}
	\frac{\ud \Gamma_{\textrm{q}}}{\ud\chi_\gamma}=\frac{\alpha m}{\sqrt{3}\pi\gamma\chi_e}
\bigg[\bigg( 2+\frac{\text{x}^2}{1+\text{x}}\bigg) K_{2/3}(\tilde{\chi}) 
 -\int_{\tilde{\chi}}^\infty \ud y\, K_{1/3}(y)\bigg], \label{constantfieldrate}
\end{equation}
where $K_\nu$ is the modified Bessel function, \RefB{$\chi_\gamma= \sqrt{k^{\prime\mu} F^{2}_{\mu\nu}k^{\prime\nu}}/m^2$} for the emitted 
photon with momentum $k'_\mu$, note that $\text{x}=\chi_\gamma/(\chi_e-\chi_\gamma)$, and 
$\tilde{\chi}= 2\text{x} /(3\chi_e)$. Although $\ud\Gamma_{\textrm{q}}/\ud\chi_{\gamma}$ diverges at small 
$\chi_{\gamma}$, the total rate of photon emission $\Gamma_{\textrm{q}}$, given by integrating (\ref{constantfieldrate}) 
over all $\chi_\gamma\in[0,\chi_e]$, is finite. (This apparent softening of the usual infra-red 
divergence in QED is explained in~\cite{Ilderton:2012qe}.)  
\Chris{A discussion of the relationship between quantum expression for the emission rate (\ref{constantfieldrate}) and the classical expression (\ref{Erikclassical}) is given in the appendix.}

In the QED simulations the electron is propagated along a classical trajectory divided into discrete time steps. After each step  $\Delta t$ the code calls a statistical subroutine to calculate the probability of photon emission and to correct the electron's momentum. The routine generates a uniform random number $r\in[0,1]$ and, if the condition $ r\leq \Gamma_\textrm{q} \Delta t$ is satisfied, an emission is deemed to occur (under the requirement $\Gamma_\textrm{q} \Delta t\ll1$).  Note that $\ud\Gamma_\textrm{q}/\ud\chi_{\gamma}$ (and $\Gamma_\textrm{q}$) are time-dependent quantities, due to the temporal variation of both the laser pulse and the electron motion. If an emission event occurs, a second uniform random number $\zeta\in[0,1]$ is generated and the photon's $\chi_{\gamma}$ (and hence its frequency) is calculated as the root of the sampling equation
\RefB{
\be
	\zeta={\Gamma_{\textrm{q}}(t)}^{-1} \int_{0}^{\chi_{\gamma}}\!\ud\chi_\gamma^\prime \frac{\ud \Gamma_{\textrm{q}}(t)}{\ud\chi_\gamma^\prime} \;.
\ee}
The photon momentum is then determined by $\chi_\gamma$ together with the assumption that at high $\gamma$ the photon is emitted in the direction of the electron's motion, just as in the classical method described above.  Finally, the momentum of the photon is subtracted from the momentum of the electron, i.e.\  the electron is recoiled, imposing the conservation law $\chi_e\to \chi_e-\chi_{\gamma}$~\cite{Ritus:1985}. The simulation then proceeds by propagating the electron (via the Lorentz equation) and the photon (on a linear trajectory) to the next time step. In this way, multiple emissions are described as sequential single photon emissions, as in (\ref{constantfieldrate}), occurring at discrete time intervals (for further information on the multi-photon calculation see Ref.~\cite{DiPiazza:2010mv} and for discussions of double emissions Refs.~\cite{Seipt:2012tn,Mackenroth:2012rb}).  This method has recently been tested against cases where we can calculate the Compton spectra analytically and found to perform extremely well \cite{Harvey:2015}.

\section{Results} 
\begin{figure}[t]
\includegraphics[width=0.49\columnwidth,clip=true,viewport=20 190 545 605]{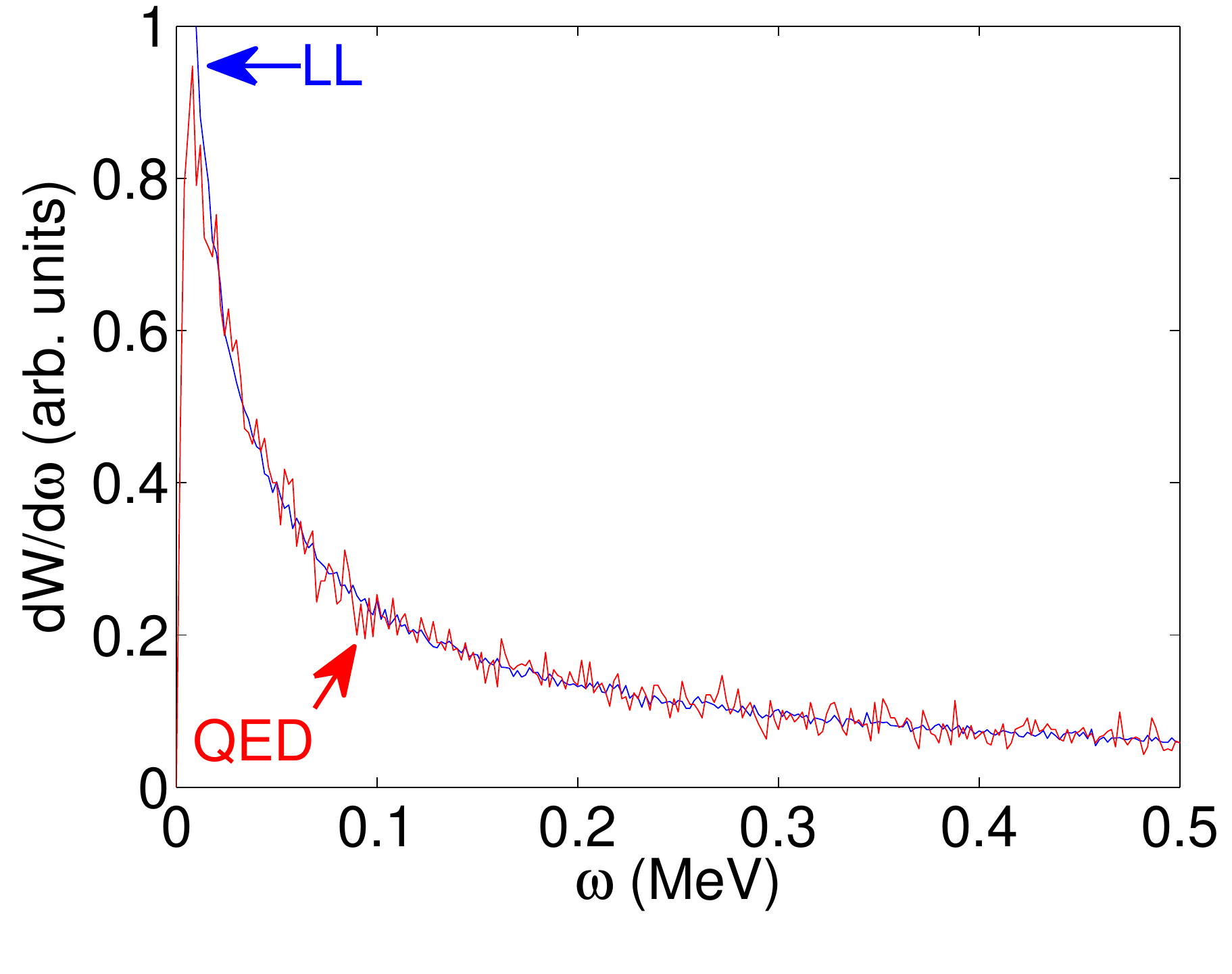}\includegraphics[width=0.49\columnwidth,clip=true,viewport=20 190 545 605]{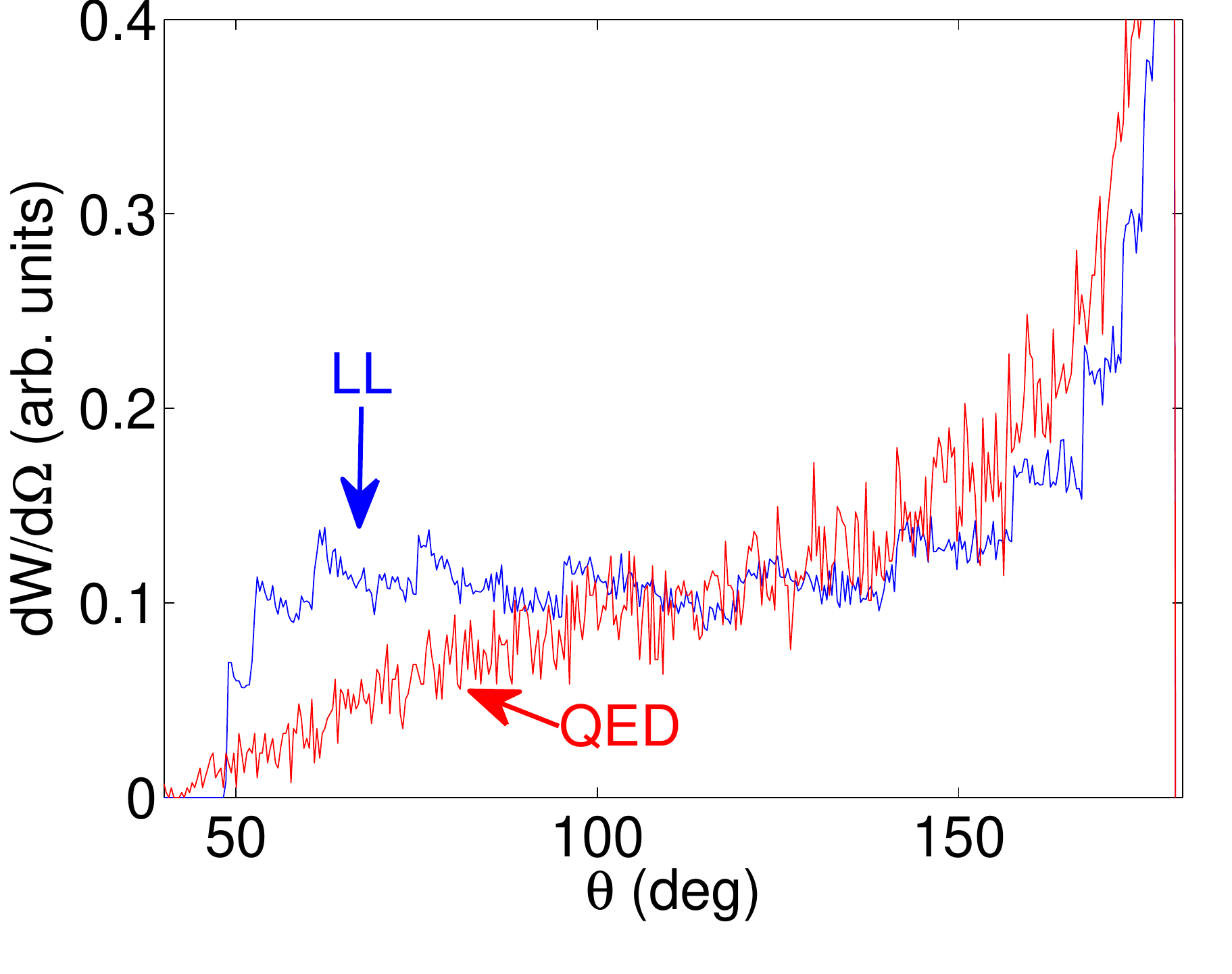}
\caption{Integrated frequency spectra (left panel) and angular spectra (right panel) for the case of an electron with initial $\gamma_0=800$ colliding with a laser of intensity $a_0=200$, $\lambda=0.8\mu$m and duration 30fs.  Blue line: classical.  Red line: QED. \label{fig:a0_200_g_800} }
\end{figure}
\subsection{Plane wave model}
We begin by considering  an electron with an initial $\gamma_0=800$ in a head on collision with a plane wave laser of peak intensity $a_0=200$, $\lambda=0.8\mu$m and duration 30fs FWHM. The emission spectra, calculated both classically and using the statistical QED routines, are plotted in Fig.~\ref{fig:a0_200_g_800}.  It can be seen that, although for these parameters there is little difference between the frequency spectra, the classical theory predicts a much stronger signal at small angles than the QED theory.
\begin{figure}
\includegraphics[width=1.0\columnwidth]{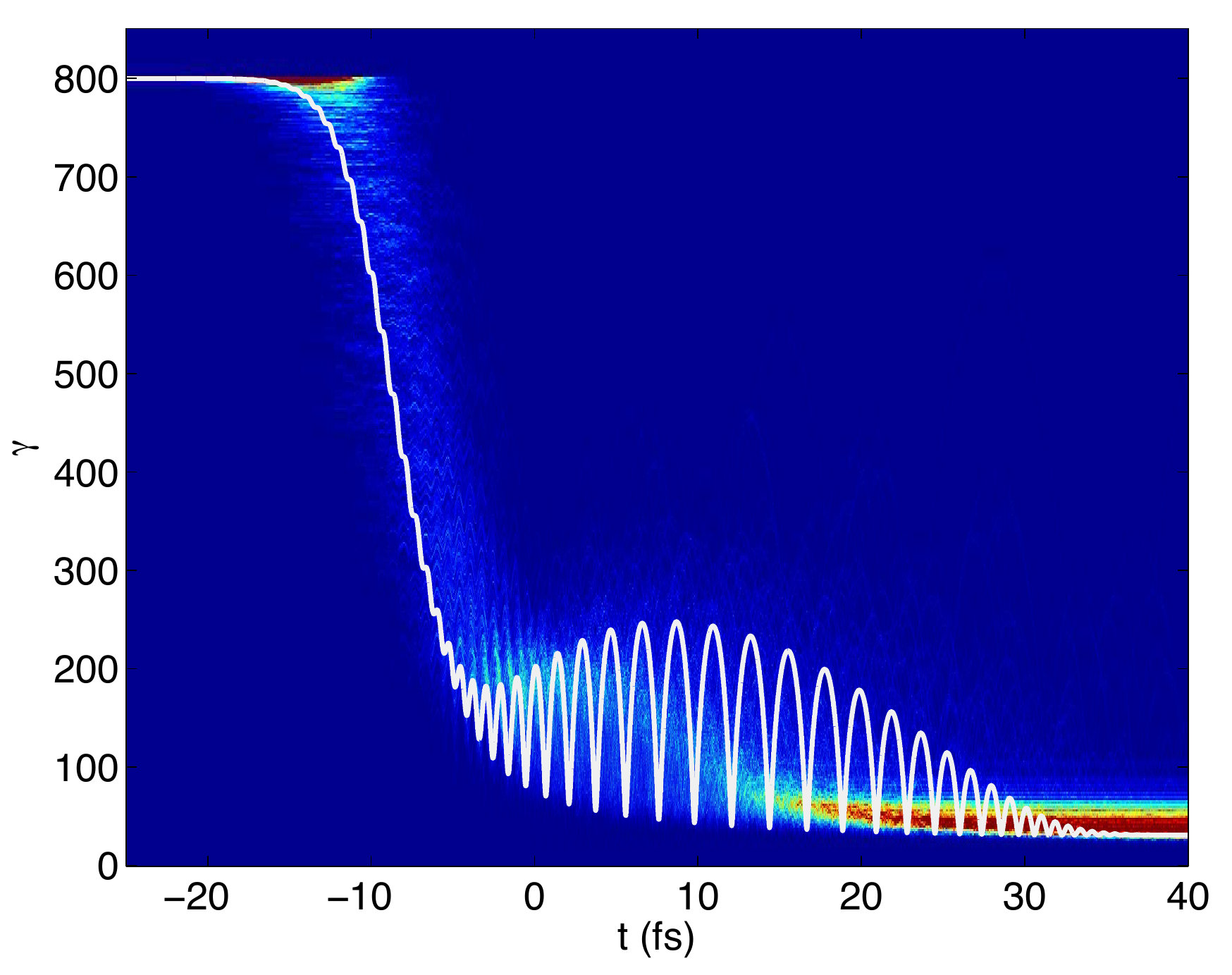}
\caption{Density plot showing how the electron $\gamma$-factor changes with time (statistical distribution generated by recording the paths of 500 QED electrons all with the same initial condition $\gamma_0=800$).  Parameters are $a_0=200$, $\lambda=0.8\mu$m and duration 30fs.  The white line shows the $\gamma$-factor for a classical electron. \label{fig:a0_200_g_800_gamma_matrix} }
\end{figure}

To understand the reason for this discrepancy we must consider how the electron energy changes with time.  In Fig.~\ref{fig:a0_200_g_800_gamma_matrix} we plot the $\gamma$-factor of the classical electron (white line).  It can be seen that the particle rapidly loses energy as it approaches the pulse focus.  This means that the angular direction of the emitted radiation will be continuously changing as the ratio of the electron $\gamma$-factor to laser $a_0$ changes with time (see Fig.~\ref{fig:centre_of_mass}).  This broadening of the angular range was first proposed in Ref.~\cite{DiPiazza:2009} as a signature of classical RR effects.
(We also note that the rapid energy loss due to RR results in a natural resistance to the high-energy high-intensity regime, and is what prevents the use of lasers for distinguishing between RR models, see e.g.~\cite{Kravets:2013}.)  Superimposed on the same plot we also show how the electron $\gamma$-factor changes in the QED case.  To generate this statistical density we ran the code 500 times, using the same initial conditions for each run.  It can be seen clearly that the classical expressions overestimate the energy loss of the particle. 
The reason for this can be understood from the fact that the QED electron only emits at discrete times, allowing it to penetrate deeper into the laser pulse before it loses energy from an emission \Chris{(an effect known as ``straggling'' \cite{Shen:1972})}, whereas the classical electron is radiating and losing energy continuously\footnote{We note also that in Refs.~\cite{Harvey:2009, Lerche:2010} it has been shown formally that, for any given frequency, the classical radiation spectrum forms an upper bound to the (single photon) QED Compton spectrum.} \cite{Burton:2014,DiPiazza:2010mv}. 
This means that, as the classical particle propagates through the pulse, its energy is going to be correspondingly lower than its QED counterpart.  From Fig.~\ref{fig:centre_of_mass} we can see that this means that the classical emissions will be at a correspondingly smaller angle than in the QED case.  This is why the QED emission spectrum dies off at a bigger angle than the classical spectrum.

\begin{figure}
\includegraphics[width=1\columnwidth]{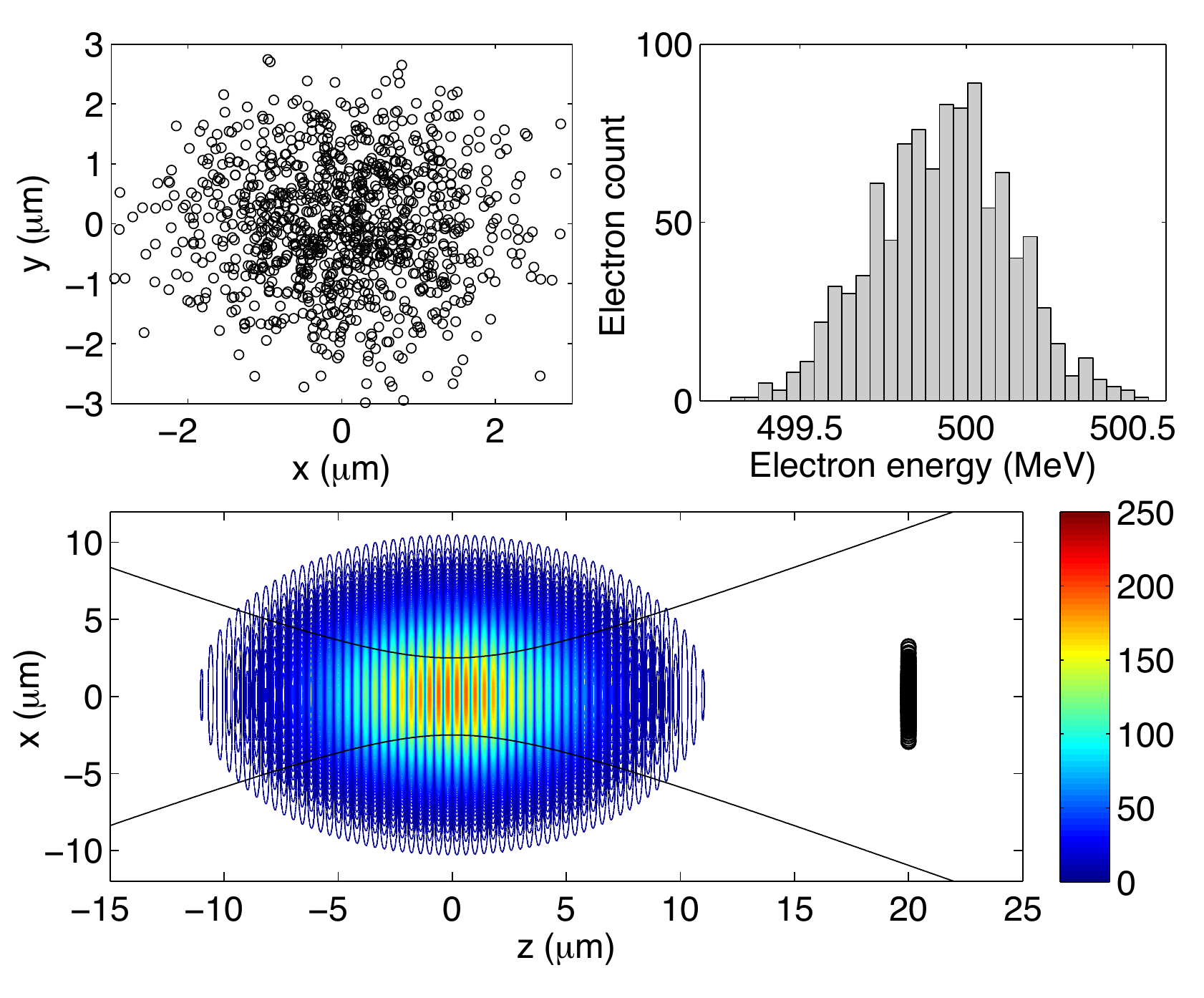}
\caption{Details of the setup for a more realistic example.  Top left panel: We consider a bunch of 1000 electrons randomly distributed in transverse space according to a Gaussian distribution of  $5\mu$m FWHM. Top right panel: initial energy distribution of the electron bunch. Bottom panel: plot showing the laser intensity (colorscale shows $a_0$) together with the electron bunch in the $z-x$ plane. \Chris{The solid black lines show how the laser waist varies as the pulse propagates (note that in the simulations the collision occurs at the origin).} \label{fig:realistic_setup} }
\end{figure}

\subsection{Realistic Example} 
To show that these results still hold in an actual experiment we consider a more realistic setup.
Instead of a plane wave field we now model our laser pulse as a 30fs duration paraxial beam, of wavelength $\lambda=0.8\mu$m, peak $a_0=250$ (equivalent to $2.64\times 10^{23}$W/cm$^2$), focussed to a waist radius of $2.5\mu$m.  
\Chris{These fields provide an accurate description of a focussed laser pulse, satisfying Maxwell's equations to the order of the paraxial expansion parameter $\theta_0=\lambda/\pi w_0\approx 0.102$, for more details see, e.g., Refs.~\cite{Davis:1979, Harvey:2012gaussian} (and for a discussion of Compton scattering in ultra-short focussed pulses see Ref.~\cite{Li:2014})}.
The parameters we have chosen are typical of what will be achievable at the ELI facility \cite{ELI}.  We model our electron source as a beam of particles initially following a Gaussian distribution in transverse space of $5\mu$m FWHM.  The particle energies average 500MeV, with a FWHM of $0.7$MeV, following the distribution shown in the top right panel of Fig.~\ref{fig:realistic_setup}.

\begin{figure}
\includegraphics[width=1.0\columnwidth]{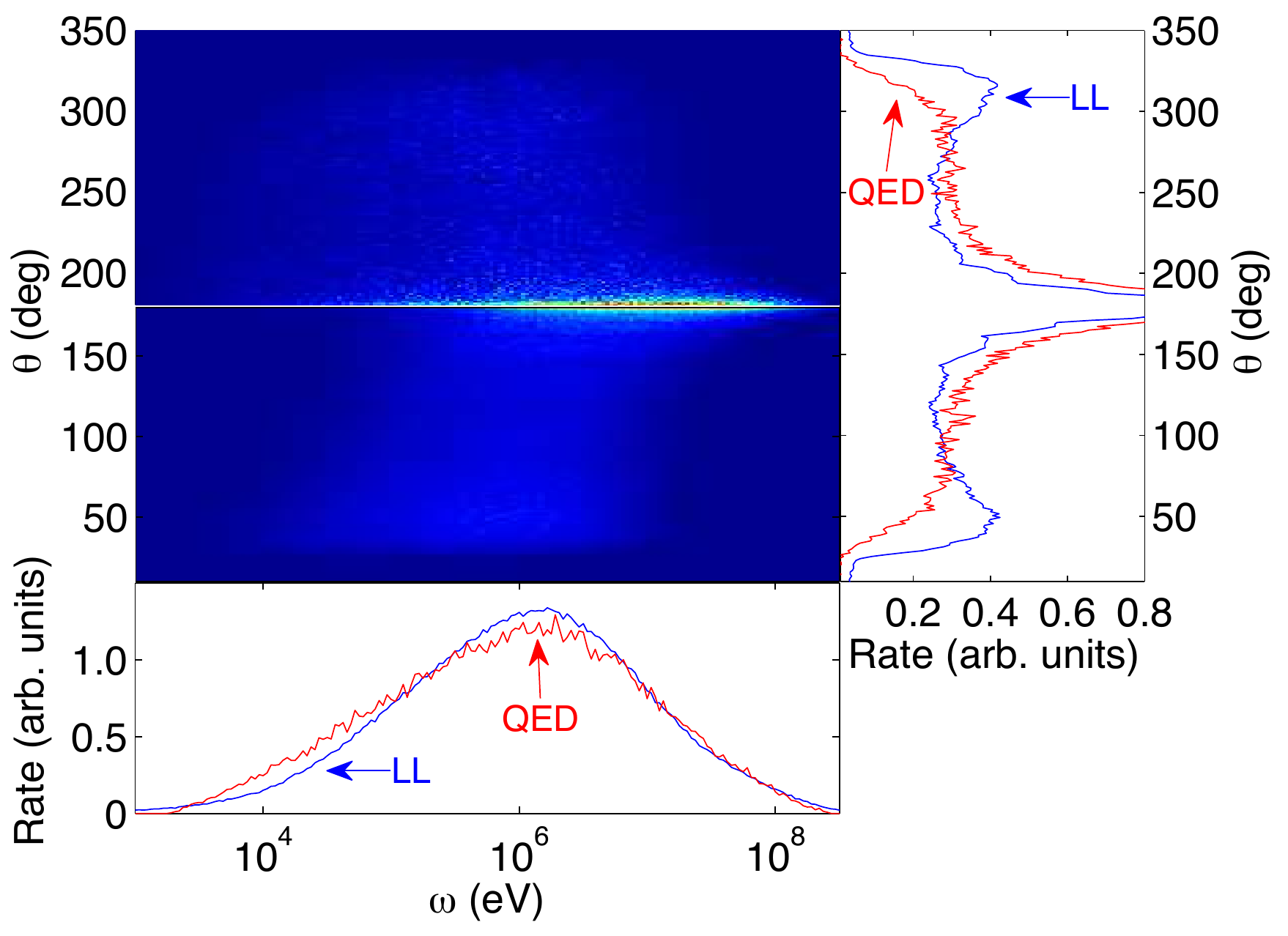}
\caption{Emission spectrum for the realistic case ($a_0=250$) described in Fig.~\ref{fig:realistic_setup}. The centre panel shows the radiation intensity as a function of frequency and angle.  This panel is split into two, the top half showing the emissions for the QED simulation and the bottom half the classical.  The right hand panel shows the total angular rate summed over all frequencies (both classical and 
QED for all angles), and the bottom panel the total frequency rate summed over all angles.  Red lines: QED.  Blue lines: classical.\label{fig:a0_250_g_500_MeV} }
\end{figure}

The resulting emission spectra of the bunch of 1000 electrons are shown in Fig.~\ref{fig:a0_250_g_500_MeV}.  Even though we are now using realistic, non-ideal parameters, the difference in angular spectra between the classical and QED models is still clearly evident. The classical theory predicts a strong radiation signal over the range $\theta\sim 10-50^\circ$, whereas the QED theory predicts only minimal radiation at these angles.  At the same time the integrated frequency spectra are still very similar for these values.    
\Chris{Hence we find that these signatures are sufficiently robust to survive realistic beam focussing effects.}

\section{Discussion} Our analysis shows a significant difference in the angular emission spectra predicted by the classical and QED theories.  The intensities we have chosen are only slightly higher than the current state-of-the-art, and well within the parameters expected at the new generation of facilities, such as ELI. 
For the regimes we have considered we find $\chi_e\sim 0.2 - 0.25$, indicating that this is an effect which becomes important before the onset of more explicit QED processes, such as pair production and runaway cascading.
The difference in angular spectra won't be visible for all sets of parameter values.  One needs a large $a_0$ in order that RR effects are significant and, additionally,  one needs for the electrons to become reflected roughly in the middle of the pulse.  
If one has $a_0\gg\gamma_0$ then the electrons will be reflected very early and thus spend a significant amount of time co-propagating with the pulse -- a regime in which the dynamics will be wholly classical.  Also, if one has $\gamma_0\gg a_0$ \RefB{the electrons} will not be reflected and the angular broadening due to RR will be small.
Nevertheless, there is a broad range of parameters where $a_0$ and $\gamma$ are of same order of magnitude and the effect will be significant. 
\Chris{(This parameter regime is found to be optimal for various studies of intense laser-particle interactions, including nonlinear Compton scattering~\cite{Harvey:2009}, classical radiation reaction~\cite{DiPiazza:2009}, and for the generation of attosecond gamma-ray pulses~\cite{Li2015:atto}.)}
 
The angular narrowing we have predicted is important for two reasons.  Firstly, it provides a clear signal of strong field QED effects, distinct from classical RR, at parameters that will soon be obtainable.  Secondly, it is an effect which will have to be taken into consideration when planning applications of Compton scattering with the new generation of ultra-intense lasers.

\acknowledgments
This research was supported by the Swedish Research Council, grants \# 2012-5644 and 2013-4248 and by the Wallenberg Foundation grant ``{\it Plasma based compact ion accelerators}". \Arkady{A.G. also acknowledges the Russian Foundation for Basic Research, project No. 15-37-21015.}

\appendix*
\section{Deriving the classical rate expression from that used in the QED case} 
\Chris{
Here we demonstrate explicitly that the expression for photon emission in the QED form (\ref{constantfieldrate}) coincides exactly with the one in the classical form (\ref{Erikclassical}) in the limit of low intensity and/or energy.  
Following the derivation of the classical synchrotron spectrum \cite{LLII} we represent Eq. (\ref{constantfieldrate}) via the first and the second synchrotron function \cite{gonoskov.pre.2015}. Firstly, we convert the rate of emission to the spectral intensity of emission, multiplying it by photon energy $\omega^\prime$ (as previously, according to the choice of units $\hbar = c = 1$)
\begin{equation}
\begin{split}
\frac{\partial I^q}{\partial\omega^\prime} =& \frac{\partial \Gamma}{\partial \chi_\gamma} \frac{ \omega^\prime}{\partial \omega / \partial \chi_\gamma} \\
=& \frac{\omega^\prime e^2}{\sqrt{3} \pi \gamma^2} \left[ \left(2 + \frac{\texttt{x}^2}{1 + \texttt{x}}\right) K_{2/3}\left(\tilde{\chi}\right)
-\int_{\tilde{\chi}}^{\infty}dy K_{1/3}\left(y\right)\right].
\end{split}
\end{equation}
Next, we apply a recursive relation of the modified Bessel function
\begin{equation}
2 \frac{\partial }{\partial t} K_{\nu}\left(t\right)= - K_{\nu - 1}\left(t\right) - K_{\nu + 1}\left(t\right)
\end{equation}
for $\nu = 2/3$ to transform the integral in the right side (note, $K_{\mu}(t) = K_{-\mu}(t)$)
\begin{equation}
\begin{split}
\frac{\partial I^q}{\partial\omega^\prime} =& \frac{\omega^\prime e^2}{\sqrt{3} \pi \gamma^2} \left[ \left(2 + \frac{\texttt{x}^2}{1 + \texttt{x}}\right) K_{2/3}\left(\tilde{\chi}\right)\right.\\
& \left. +\int_{\tilde{\chi}}^{\infty}dy \left(K_{5/3}\left(y\right) + 2 \frac{\partial }{\partial y} K_{2/3}\left(y\right)  \right)\right].
\end{split}
\end{equation}
Taking into account that $\lim_{y \to \infty} K_{5/3}(y)= 0$ we obtain
\begin{equation}
\frac{\partial I^q}{\partial\omega^\prime} = \frac{\sqrt{3}\omega^\prime e^2}{2 \pi \gamma^2} \frac{\chi_e \left(\chi_e -\chi_{\gamma}\right)}{\chi_{\gamma}}\left[\frac{\texttt{x}^2}{1 + \texttt{x}}F_2\left(\tilde{\chi}\right) + F_1\left(\tilde{\chi}\right)\right],
\end{equation}
where $F_2(\xi) = \xi K_{2/3}(\xi)$ is the second synchrotron function. In terms of the \textit{effective magnetic field} $H_{\textrm{eff}}$ we can express
\begin{equation}
\chi_{e} = \gamma \frac {H_{\textrm{eff}}}{E_{\textrm{cr}}},  \quad  \chi_{\gamma} = \frac{\omega^\prime}{m}\frac {H_{\textrm{eff}}}{E_{\textrm{cr}}}.
\end{equation}
Using the ratio of the photon energy to the initial electron energy $d = \omega^\prime / \left(m \gamma\right) = \chi_{\gamma} / \chi_e$ as a parameter we obtain
\begin{equation}
\begin{split}
\frac{\partial I^q}{\partial\omega^\prime} =& \frac{\sqrt{3}}{2 \pi} \frac{e^3 H_{\textrm{eff}}}{m} \left(1 - d\right) \times \\
&\left[\frac{d^2}{1 - d}F_2\left(\frac{2}{3\chi_e}\frac{d}{1-d}\right)   + F_1\left(\frac{2}{3\chi_e}\frac{d}{1-d}\right)\right].
\end{split}
\end{equation}
This form of the expression makes it possible to see that in the case where the photon has much lower energy than the electron ($d \ll 1$) the expression tends exactly to the classical one (\ref{Erikclassical}).}

%

\end{document}